\documentclass[pre,twocolumn,superscriptaddress,showpacs,pdfoutput=1]{revtex4-1}
\usepackage{graphicx}
\usepackage{epsfig}
\usepackage{amssymb,amsmath,amsfonts,hyperref}
\usepackage{wasysym}
\usepackage{bm}
\usepackage{latexsym}
\usepackage{eucal}
\usepackage{float}
\usepackage{verbatim}
\usepackage[normalem]{ulem}
\usepackage{epstopdf}
\usepackage{color}
\usepackage{braket}

\newcommand{\be}{\begin{equation}}
\newcommand{\ee}{\end{equation}}

\newcommand{\bea}{\begin{eqnarray}}
\newcommand{\eea}{\end{eqnarray}}

\begin{document}

\title{Hilbert Space Fragmentation and Ashkin-Teller Criticality\\ 
in Fluctuation Coupled Ising Models}

\author{Pranay Patil}
\affiliation{Department of Physics, Boston University, 590 Commonwealth Avenue, Boston, Massachusetts 02215, USA}

\author{Anders W. Sandvik}
\affiliation{Department of Physics, Boston University, 590 Commonwealth Avenue, Boston, Massachusetts 02215, USA}
\affiliation{Beijing National Laboratory for Condensed Matter Physics and Institute of Physics, Chinese Academy of Sciences, Beijing 100190, China}

\begin{abstract}
We discuss the effects of exponential fragmentation of the Hilbert space on
phase transitions in the context of coupled ferromagnetic Ising models
in arbitrary dimension with special emphasis on the one dimensional case.
We show that the dynamics generated by quantum fluctuations is bounded within
spatial partitions of the system and weak mixing of these partitions caused
by global transverse fields leads to a zero temperature phase with ordering in 
the local product of both Ising copies but no long range order in either 
species. This leads to a natural connection with 
the Ashkin-Teller universality class for
general lattices. We confirm this for the periodic chain using quantum Monte
Carlo simulations. We also point out that our treatment provides an explanation
for pseudo-first order behavior seen in the Binder cumulants of the classical 
frustrated $J_1-J_2$ Ising model and the $q=4$ Potts model in 2D.
\end{abstract}

\maketitle


\section{Introduction}

The nature of quantum phase transitions has generated a large amount of interest
in the context of magnetic systems. Some of the important fields in which the
physics at a quantum phase transition plays an essential role are order to
order transitions with exotic emergent symmetries
\cite{Senthil,Nahum,NahumX,Bowen}, determining the ability of quantum
annealing to solve computational problems \cite{qa1,qa2,qa3}, 
and the understanding of field 
theoretic frameworks to describe low energy physics of discrete models
\cite{IsingCFT,PottsFT,ClockFT}.
Crucial to these topics is the structure of low energy excitations at 
critical points, especially those which have spatial restrictions such as
fractons \cite{FracRev,FracCham}. Recently these restricted dynamics have
been seen as a consequence of spatial ``fragmentation'' of the
Hilbert space \cite{ETH1,ETH2}. Fragmentation describes the consequence of
block diagonalization of a Hilbert space into an exponentially large number 
of sectors, with spatial structure corresponding to the states making up a
sector. A similar phenomenon has been observed in quantum dimer models as well \cite{Sha}, 
although a spatial pattern corresponding to sectors has not been identified in this case.

In this article, we address another more general manifestation of the 
phenomenon of Hilbert space fragmentation
by introducing a simple model whose Hilbert space breaks into an 
exponentially large number of sectors, each of which has interesting spatial 
patterns which limit the growth of the correlation length. 
We draw connections between
the nature of this fragmentation and the partitions of natural numbers,
and discuss in the context of this model
the effects of the underlying lattice it is set on in terms of its
percolation properties and the structure of excitations they lead to. 
We study the nature of
eigenstates and energies for the 1D case 
and briefly discuss our expectations
from this model when it is placed in contact with a thermal bath. We then turn
to the effects of adding symmetry breaking quantum perturbations at zero
temperature, where we find behavior suggestive of a phase with partial ordering.
We draw a comparison with the Ashkin-Teller model \cite{ATintro}, 
where a similar phase is seen, and argue for a complete mapping between our 
model and the Ashkin-Teller model. 
We quantitatively check this mapping for the 1D case 
using quantum Monte Carlo simulation,
and find consistency with the range of continuously
varying exponents already known to exist for the Ashkin-Teller universality 
class \cite{Cardy,ATsource}.
We also point out that the partially ordered phase provides an explanation for
pseudo-first order behavior observed in the Binder cumulants at some continuous
phase transitions, e.g. 2D $q=4$ Potts and $J_1-J_2$ frustrated Ising 
models \cite{JinSan}.

The outline of the paper is as follows: In Sec.~\ref{Sec2}, we present the
model, describe the fragmented Hilbert space structure, and point out the
few general constraints required to get this feature. We also present a 
detailed study of the energy spectrum for a periodic chain. 
In Sec.~\ref{Sec3}, we
incorporate the perturbation which takes the systems away from the fragmented
Hilbert space structure and briefly discuss the expected effect on 
dynamics. This is followed by a description of the Ashkin-Teller universality
class along with a general mapping to our system, which we check in detail for 
the 1D system through numerical results. In Sec.~\ref{Sec4}, we describe 
pseudo-first order behavior and the role of the partially ordered phase
in generating a behavior similar to the 4-state Potts and other related models.
In Sec.~\ref{Sec5}, we conclude with a brief summary and discussion.


\section{\label{Sec2}Fluctuation Coupled Ising Models and Fragmentation}

We will introduce the fragmentation of the Hilbert space and its consequences
in the context of a coupled Ising model made out of two Ising species,
$\sigma$ and $\tau$, with the following Hamiltonian:
\be\label{oham}
H=-\frac{s}{2}\sum_{\langle i,j\rangle}\big(\sigma^z_i\sigma^z_j+
\tau^z_i\tau^z_j \big)-(1-s)\sum_i\sigma^x_i\tau^x_i.
\ee
Here, $\langle i,j\rangle$ refers to nearest neighbors and $s$ is the 
tuning parameter
used to drive the ground state from a paramagnet ($s=0$) to a ferromagnet
($s=1$). This model can also be written using just a single species ($\sigma$)
which lives on a larger lattice and comprises of two copies of the 
original lattice connected in a bilayer fashion. 

The Hamiltonian described in Eq.~\eqref{oham}
possesses a local conserved quantity, $\sigma^z_i\tau^z_i$, associated with 
each site of the lattice. This is due to the particular form of the quantum
fluctuation, $\sigma^x_i\tau^x_i$, which commutes with $\sigma^z_i\tau^z_i$. 
As $\sigma^z_i\tau^z_i$ is a
conserved quantity, it takes well defined values, which are $+1$ and $-1$
in this case. For a lattice with $N$ sites, the set 
$\{\sigma^z_i\tau^z_i\}^N_{i=0}$ can take $2^N$ values, implying that the
Hamiltonian can be broken into $2^N$ blocks. An example of one such block is
shown in Fig.~\ref{fa2d} for a square lattice, where $+/-$ denote the value
of $\sigma^z_i\tau^z_i$ at each site. In terms of the individual $\sigma$ and
$\tau$ spins, we shall use $0$ to denote the state
$\sigma^z(\tau^z)=-1$ and $1$ to denote $\sigma^z(\tau^z)=+1$. Now the
four possible states in the $z$-basis on a single 
site are $\{00,01,10,11\}$ where the first number denotes the state of the 
$\sigma$ spin (0 or 1) and the second denotes the $\tau$ spin. The constraint
of a particular value for $\sigma^z_i\tau^z_i$ reduces the number of basis
states at site $i$ to two. This implies each block is a $2^N\times 2^N$ matrix.

For the rest of our analysis, we will assume that we have block diagonalized
the Hamiltonian in Eq.~\eqref{oham}, and that any quantum state of the system
belongs only to one of these blocks.
In addition to the block structure due to the conserved quantities described
above, there is a further fragmentation within each block. This is developed
below for the cases of an arbitrary lattice and a periodic chain.

\subsection{Arbitrary Lattice}

We begin by considering a system which lives 
on an arbitrary lattice or dimensionality and examine the spin degrees of
freedom. We show below that for nearest neighboring sites $i$ and $j$, 
if the state of the system belongs to a block such that 
$\sigma^z_i\tau^z_i\neq\sigma^z_j\tau^z_j$, then this pair of sites can be
considered to be non-interacting.

\begin{figure}[t]
\includegraphics[width=0.4\hsize]{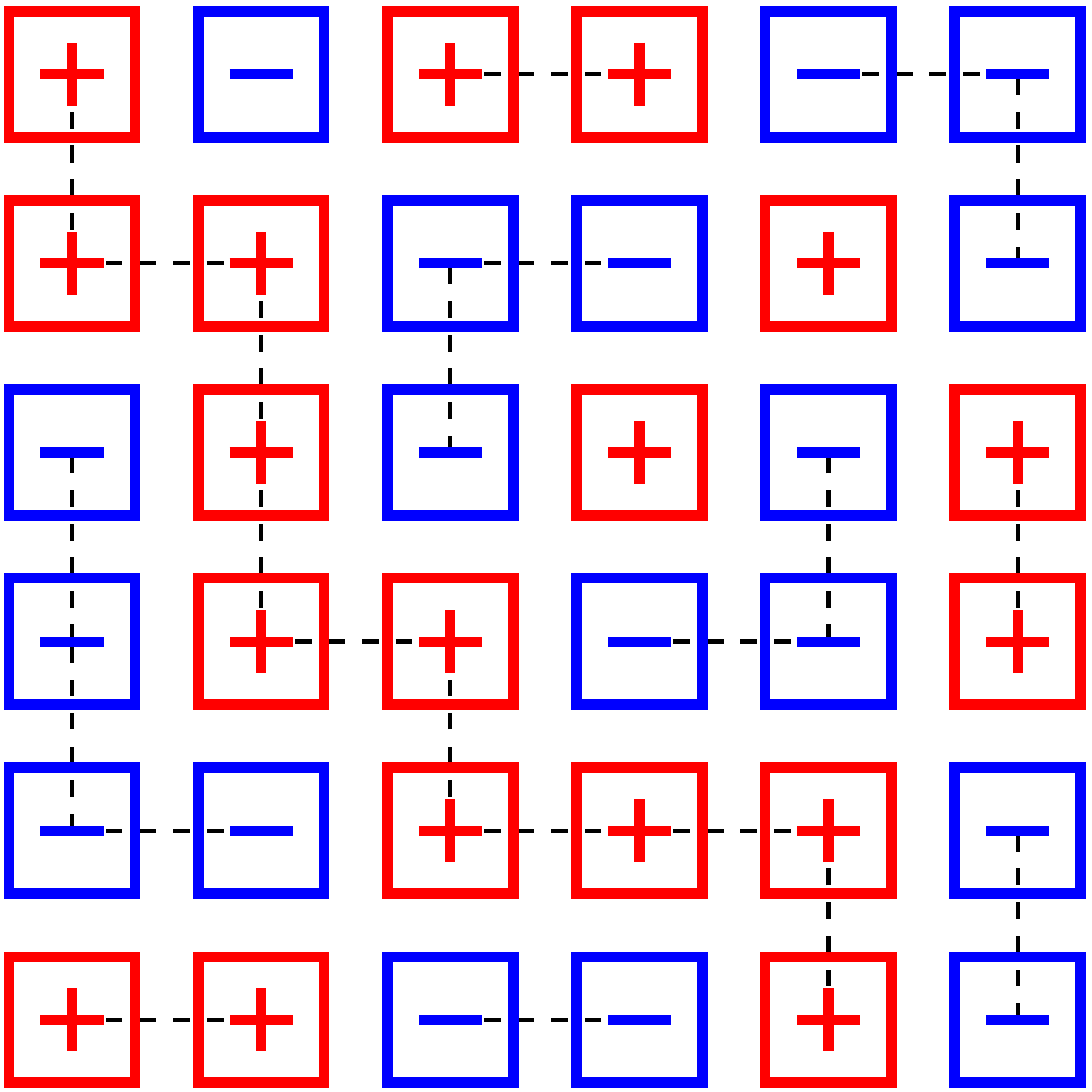}
\caption{A fragment arrangement for the square lattice with interacting sites
connected by dotted lines. The +'s and -'s denote the value of 
$\sigma^z_i\tau^z_i$ at each site.}
\label{fa2d}
\end{figure}

First consider $\sigma^z_i\tau^z_i=\sigma^z_j\tau^z_j=+1$, with the state $00$ 
on site $i$ and the state $11$ on site $j$. The energy cost of 
such an arrangement due to the classical term $-(\sigma^z_i\sigma^z_j+
\tau^z_i\tau^z_j)$ is -2 (in units of the ferromagnetic coupling). This can
be maximized by switching $11\to 00$ on site $j$, which is allowed by
$\sigma^z_j\tau^z_j=+1$. If the constraint on site $j$ is instead chosen to
be $\sigma^z_j\tau^z_j=-1$, then site $j$ hosts either $01$ or $10$, and the
energy associated with the arrangement of states on $i$ and $j$ will always
be 0, as one of the bonds (either $\sigma\sigma$ or $\tau\tau$) is always
broken while the other is always satisfied. This implies that, from energy
considerations, states $01$ and $10$ are equivalent if site $i$ hosts $00$.
It follows that, if $\sigma^z_i\tau^z_i=\sigma^z_j\tau^z_j$, 
then they have an Ising bond of strength $s$ between them, else, they are
non-interacting. If we now consider a typical set of values 
of $\{\sigma^z_i\tau^z_i\}^N_{i=0}$ (corresponding to a block)
as shown in Fig.~\ref{fa2d} for a 2D lattice, we see that the 
system has essentially broken into
several smaller Ising models which co-exist on the lattice. 
We call each of these smaller Ising models a fragment.
For simple regular lattices, such
as a 1D chain, square or cubic lattice with periodic conditions, the fragment
arrangements can be related to the partitions of natural numbers
\cite{Part1,Part2}. This
connection will be later illustrated using a periodic chain. 
It is worth noting here that this argument applies also for a more general
classical term of the form $f(\sigma^z_0,\tau^z_0,...,\sigma^z_N,\tau^z_N)$,
which is symmetric under the exchange of the Ising species, $\sigma\to\tau$.
As this function must be symmetric for all spin configurations, the constraint
would need to be satisfied for each bond, and the above argument
would be valid. There may be other convoluted ways to satisfy the symmetry 
requirement for certain complicated functions.

The phenomenon described above is similar to the fragmentation discussed
in recent work
in the context of the eigenstate thermalization hypothesis (ETH)
\cite{ETH1,ETH2} and has also been studied in 
disordered Floquet circuits composed of Clifford
gates \cite{ACCL}. This phenomenon has also been observed numerically in 
quantum dimer models with restricted dynamics, but a similar real space 
geometric way to understand the same has not been identified in that 
context \cite{Sha}.

One of the key features of the 
fragmentation of real space into components is that the correlation length
in a particular block is bounded by the spatial extent of the largest
clusters in the corresponding fragment arrangement. 
This feature depends crucially on the restricted dynamics generated by 
$\sigma^x_i\tau^x_i$ and the classical term allowing a degeneracy in energies.
If the classical term were to be augmented by adding an interaction of the
form $-\sigma^z_i\tau^z_j$ which breaks the $\sigma\to\tau$ symmetry, 
the non-interacting nature would be lost as the
state $00$ on site $i$ would now prefer $10$ on site $j$ over $01$.
Due to this term, each spin species has a global pattern specific to which
block the state belongs to, and fluctuations would occur around this
pattern in the large $s$ limit. The maximal correlation length in every block
grows to the system size in the presence of such a term, although the 
details of this growth depend
on the structure of the particular block. These arguments illustrate that, 
although the quantum term determines the block structure, interacting units 
within a block may be controlled by the choice of classical terms. Careful 
choice of tuning parameters can also create a scenario where there are two 
length scales, one associated with the growth of correlation within a component
and the other with the growth across components. If we were to
require that the symmetry in $\sigma\to\tau$ be maintained, an additional
term would have to be added to ensure that the state $00$ does not
favor one of $01$ or $10$, and the physics would again be the same as
the Hamiltonian in Eq.~\eqref{oham}.

\begin{figure}[t]
\includegraphics[width=0.85\hsize]{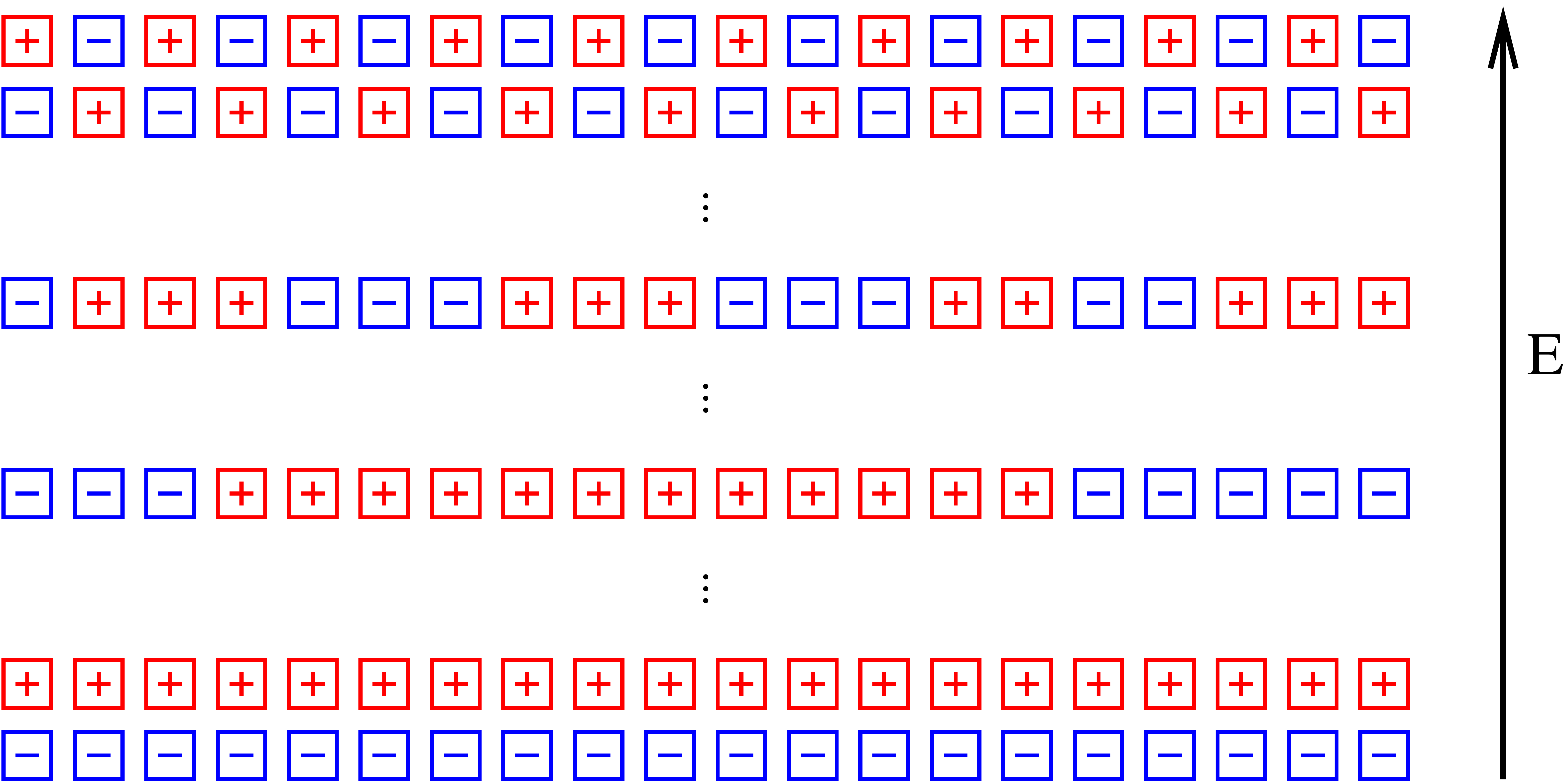}
\caption{Fragment arrangements sorted by energy with reference blocks making up
the lowest energy region and energy increasing bottom to top.} 
\label{faxp}
\end{figure}

As a particular block in the Hamiltonian can be
thought of as a configuration where each site is assigned either $+1$
or $-1$ with probability $1/2$, it can also be written in terms
of a percolation problem where a particular site is occupied or left empty. 
If the percolation threshold for the particular lattice is below $1/2$, most
blocks will have a giant fragment and this may have consequences on
the correlation length as far above the percolation threshold, almost all
blocks will have diverging length scales, leading to a
continuous transition. The universality class for the transition may relate
to those of diluted Ising models above the percolation threshold,
which have been studied in the context of thermal and quantum phase 
transitions \cite{Perc1, Perc2, Perc3}.
We are unable to study this aspect of the problem in the context of the 
periodic chain as the percolation threshold in 1D is unity, which means 
none of the fragment arrangements percolate except the two blocks which
correspond to all sites having the same value for $\sigma^z_i\tau^z_i$.
It is also noteworthy that, due to this fragmentation, the dynamics
generated by $\sigma^x_i\tau^x_i$ reduces to the dynamics within the fragments
and no inter-fragment correlations are introduced.
In 2D and higher, the fragments would in general represent non-integrable
systems which thermalize within their boundaries but not with the entire
system. This implies that the system would not reach a thermal distribution
and would not exhibit characteristics such as volume law entanglement entropy 
which would be expected of thermalizing systems.


\subsection{Periodic Chain}

Now we specialize to the case of a periodic chain with $L$ sites 
and consider the eigenstates and
eigenenergies for various $s$. In the ferromagnetic limit ($s=1$), the lowest
energy belongs to two blocks, one with $\sigma^z_i\tau^z_i=+1$ for all $i$, and 
the other $\sigma^z_i\tau^z_i=-1$ for all $i$.
We shall refer to these two blocks as the reference 
blocks for this model, as they are the easiest to analyze and map exactly
to the simple transverse field Ising chain.
These blocks have $L$ activated bonds, whereas
all other blocks have at least one pair of nearest neighbor sites which are
non-interacting, leading to loss of the energy which could potentially be gained
from that Ising bond. The first excited level in the ferromagnetic limit
is made out of all blocks which have two fragments each,
one with all $\sigma^z_i\tau^z_i=+1$ and the other with all 
$\sigma^z_i\tau^z_i=-1$, as these
arrangements have $L-2$ activated bonds. In the limit of $s=0$, all the Ising
bonds are switched off and all $2^L$ blocks are degenerate. The fragment 
arrangement of any block can be seen as a sum of independent Ising chains of 
various lengths.
As the energy density of a longer Ising chain is always larger in magnitude
than a shorter chain for all $s$ other than $s=0$ and $s=1$, the reference
blocks, which comprise a single periodic chain of length $L$, 
always form the lowest energy manifold. 
A schematic energy spectrum classification for general $s$ can be seen in 
Fig~\ref{faxp}.

As each block can be made up of many smaller chains, the energy spectrum of
this Hamiltonian hosts a large amount of degeneracy. This can be understood
by recognizing that many blocks share the same number and sizes of chains,
and each block carries a different ordering of the chains. As the energy of
a block is simply the sum of the energies of individual chains, the arrangement
of chains that makes up a particular block does not play any role in calculating
the energy; only the number and sizes of chains control the energy. Following
this line of thought, we can now map our energy spectrum to the
partitions of the natural number $L$, where each partition is defined as a set
of smaller pieces whose lengths sum to $L$. For example the partitions
for $L=4$ are:
\begin{equation}
\begin{split}
4=3+1=2+2=2+1+1=1+1+1+1
\end{split}
\end{equation}
It was shown \cite{Partm} that the number of partitions $p(L)$
of a natural number $L$ asymptotically behaves as $\log p(L)\approx C\sqrt{L}$
with $C=\pi\sqrt{2/3}$. Due to periodic boundary conditions, the only allowed
partitions for the chain arrangements are those which have an even number of
chains. It follows that the number of energy levels in addition to the reference
level are the number of even partitions of the number $L$. We observe that this
number quickly approaches half the asymptotic value for $L\geq 10$. If we now
consider a particular block and study the growth of its correlation length
as we change $s$ from the paramagnetic regime to the ferromagnetic regime,
we would find that the correlation length grows until it reaches an upper
bound which must be smaller than the largest chain in the partition 
corresponding to that block. If the largest chain is much smaller than system
size the ground state of this block can never develop long range order.
The statistics of different blocks along with their energies now control
how much they contribute to the ground state of the total system in the
presence of a temperature or symmetry breaking quantum fluctuation which allows 
them to mix.

\begin{figure}[t]
\includegraphics[width=\hsize]{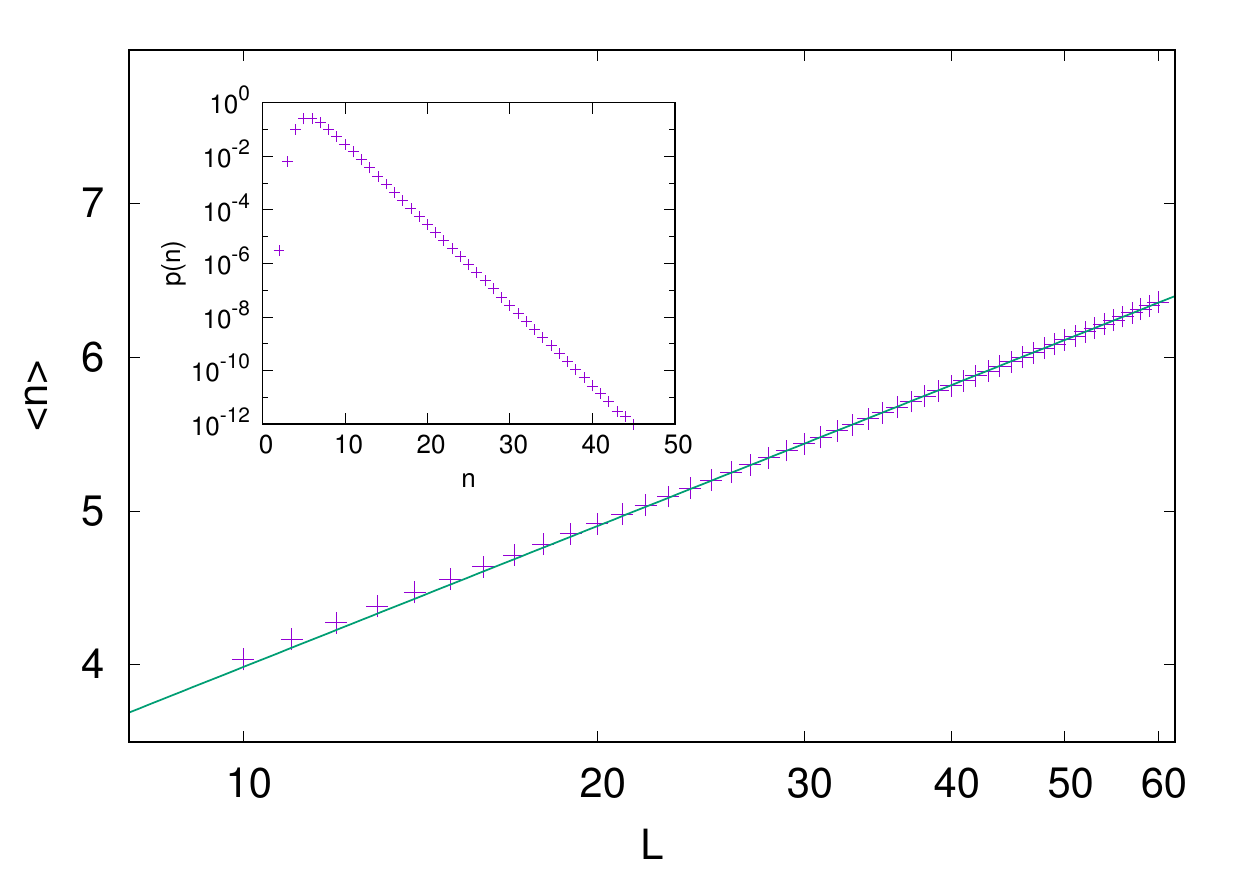}
\caption{Average size of the largest fragment in a block plotted 
as a function of the system size $L$ for a periodic chain and fit to the form 
$\braket{n}\approx a+b\log(L)$ with $b=1.32(1)$. Inset: The probability 
distribution of the size of the largest fragment, 
which shows an exponential tail.}
\label{favg}
\end{figure}

An added level of complexity is brought in by observing that each partition
of the number $L$ corresponds to a different number of chain arrangements, i.e,
blocks. The combinatorial factor related to this can be calculated using the
following arguments.
A particular arrangement of the chains in a partition can correspond to
only two arrangements for the signs of $\sigma^z_i\tau^z_i$, 
as the moment a value is chosen for a
chain, all others must be chosen in accordance with it to ensure the condition
that the pieces are non-interacting. 
Taking this into account along with
all the permutations of a particular partition and the translation invariance
of the system, we conclude that the number of blocks $b(p)$ that correspond to 
that particular partition $p$ of the system size $L$ with an even number of 
chains $N$, is
\be\label{pcomb}
b(p)=2L\frac{(N-1)!}{k_1!k_2!...k_L!},
\ee
where there $k_i$ is the number of pieces of length $i$, and we use $0!=1$. 
For example, the partition of system size $L=20$ given by $4+3+3+2+2+2+1+1+1+1$
has $N=10$ with $k_1=4$, $k_2=3$, $k_3=2$, $k_4=1$, and $k_i=0$ 
for all other $i=5,...,20$.
We have checked Eq.~\eqref{pcomb} against exact enumeration. Although
the average size of the largest chain in a partition 
where all partitions are sampled with equal weight goes as $O(\sqrt{L}\log(L))$
\cite{Partm}, we find numerically (Fig.~\ref{favg}) that the added suppression 
caused by the factor in Eq.~\eqref{pcomb} reduces this to $O(\log(L))$.

We find numerically that the average
size, $\braket{n}$, of the largest chain in the fragment 
arrangement corresponding to a random block follows
the relation $\braket{n}=a+b\log(L)$, as seen in Fig~\ref{favg} 
for $L=20,..,60$. We also study the
probability distribution of the size of the largest chain in a random block
chosen with uniform probability and find exponential tails for $p(n)$
for $n>\braket{n}$ (shown in inset of Fig~\ref{favg} 
for a 60 site chain). This suggests that,
if the system is allowed to choose a block at random, the largest chain in
the chosen block will be much smaller than system size with a probability 
$\to 1$, thus leading to a severe limitation on the growth of correlation
length.

\begin{figure}[t]
\includegraphics[width=\hsize]{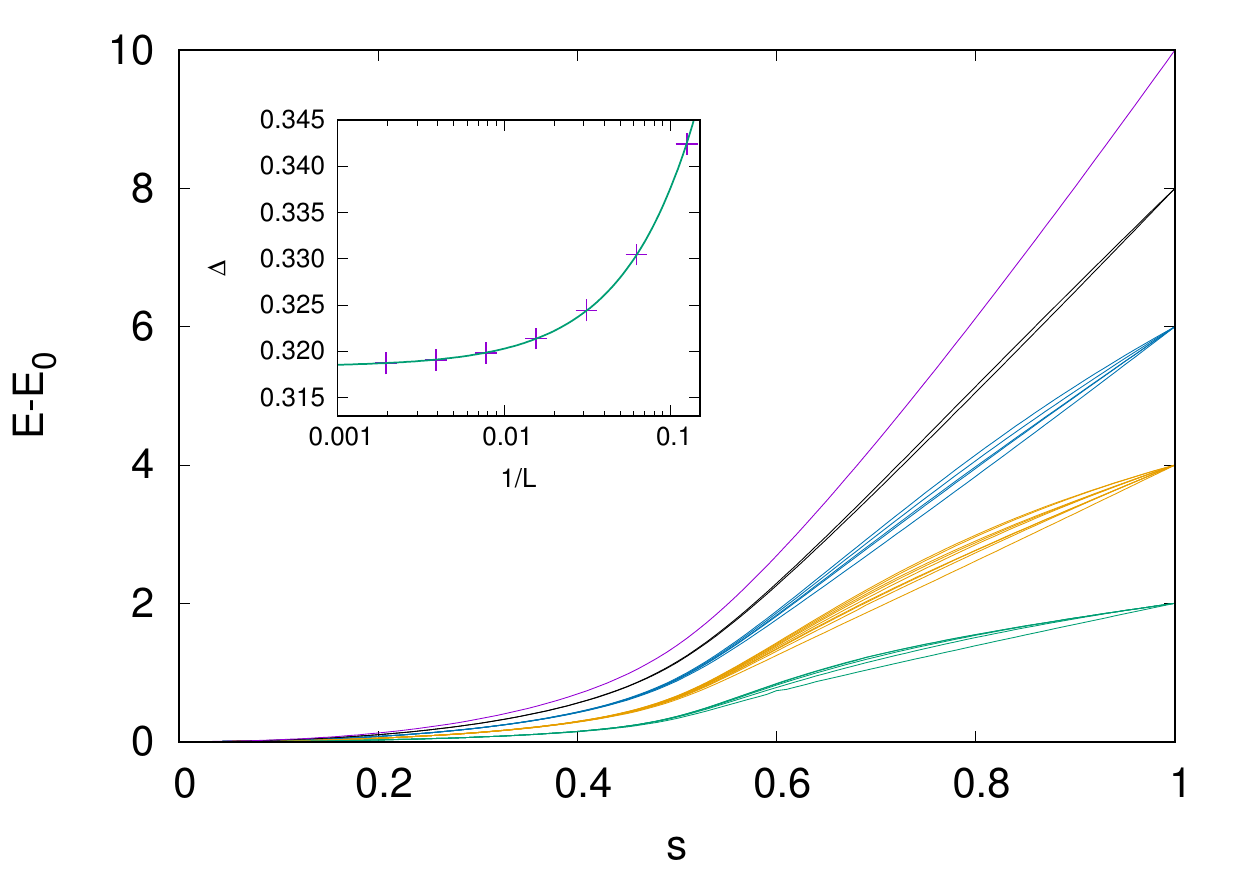}
\caption{Energy levels for the Hamiltonian defined in Eq.~\eqref{oham} for
system size $L=10$, 
as a function of tuning parameter $s$, seen to
converge in $s=0$ limit and approach the spectrum of the classical
Ising chain in $s=1$ limit. Inset: The minimum gap at $s=0.5$ as a function
of $1/L$; extrapolating to $1/L=0$ gives a minimum gap of 0.31835(2).}
\label{flan}
\end{figure}

The probability distribution with which the system samples different blocks
depends on the terms connecting different blocks and the relative ground state 
energies of different blocks. 
As discussed above, the ground state of the entire system 
is always made out of the two
blocks which have the same value of $\sigma^z_i\tau^z_i$ on all sites. 
The opposite limit
is again made up of just two blocks, which are the blocks where all odd sites
have the same value of $\sigma^z_i\tau^z_i$, and the even sites have the 
opposite value (Fig~\ref{faxp}). Each of these breaks into $L$
disconnected spins, as no nearest neighbor spins have a ferromagnetic bond
between them. This implies that every spin is polarized in the $\pm x$-direction
due to the $\sigma^x\tau^x$ term with an energy of $-s$, making the total energy
of the state $-Ls$. We can assume that the ground state energy for the
reference blocks can be written as $-L\epsilon(L,s)$ where $\epsilon(L,s)$ is
the energy density for a periodic chain of length $L$ at tuning parameter value
$s$. These two extremes set the range of energies which can be occupied by all
other blocks. Another general trend to be expected from the lowering of the 
energy due to larger system size would be to have partitions with the largest 
chains having ground states which occupy lower energy levels (Fig.~\ref{faxp}). 
As we have seen from the distribution of partitions,
these levels would contain a relatively small number of ground states as the 
parent blocks must contain long chains. 
Also, all the energies must converge in the $s=0$
limit, as the ferromagnetic term switches off, leaving all blocks equivalent
in energy. 

The above arguments give us a fair idea of the energy level diagram for the 
ground states in all the blocks.
We present a detailed study of the $L=10$ case in Fig.~\ref{flan},
obtained using Lanczos
diagonalization, which captures the essential features. An important region of
the energy level diagram is $s\approx 0.5$, as the simple Ising chain undergoes 
a continuous quantum phase transition at this point. 
In an Ising chain, the correlation length
grows continuously with increasing $s$ for $s<0.5$ and at the transition the
correlation length reaches the system size. 
If the gap to a large number of blocks 
vanishes at this point, the correlation length would acquire 
large contributions 
from the other blocks in the presence of arbitrarily small coupling across 
blocks, which would lead to a capping on the correlation length. As the
gap must once again open in the ferromagnetic regime, the system will drop back
into the fully polarized state with large correlation length. This mechanism
can create a jump in the correlation length, which is a hallmark of a first
order phase transition. This is a heuristic argument which does not take into
account the nature of the coupling to other blocks.
Using the Jordan Wigner transformation to map the Ising chains to
non-interacting fermions \cite{YoungReiger}, we find 
that the gap at $s=0.5$ indeed converges to a non-zero value with increasing 
size. This is shown in the inset of Fig~\ref{flan} for system sizes up 
to $L=512$, along with a finite size extrapolation, 
leading to a gap of 0.31835(2) in the thermodynamic
limit. This is expected for higher dimensions as well, as the
lowest block above the ground state block must necessarily have at least one
missing ferromagnetic bond which contributes a finite amount to the energy.
Our analysis also showed that the first state above the ground state of the
reference blocks belongs to a block which breaks into a fragment which has a
single site and a fragment which has $L-1$ sites. At $s=0.5$, blocks with this
type of structure have the minimum energy amongst all blocks with only two
fragments.

\subsection{Fluctuations between blocks and block mixing}

Here we discuss the effects of block mixing for arbitrary lattices with $N$
sites. One of the easier ways to allow the system 
to access all possible blocks would be
to couple it to a thermal bath which provides an inverse temperature $\beta$. 
Assuming that the ground state energies of all the blocks is $O(N)$ 
(an example of which we see in the energy level diagram in Fig.~\ref{flan}), 
the contribution of the blocks with relatively small fragments or ``restricted''
blocks ($Z_{r}$) in the partition function is $Z_r=e^{-\beta E}D_r$,
where $D_r$ is the degeneracy of the blocks. As we have seen that this 
degeneracy$\to 2^N$ and $E\propto N$, a finite $\beta$ will not necessarily 
suppress these
levels, and there can exist a range of temperatures where these levels can 
mediate a transition with limited correlation length, i.e, a first order
transition. Finite temperature would allow thermal fluctuations which can
jump across blocks and in this way wash out the block structure as well.
This cannot be studied in our analysis of the 1D chain as it is
known that any non-zero temperature leads to disorder in the Ising chain
and the phase transition is thus completely washed out.
For higher dimensional systems this mechanism can
lead to interesting crossover physics between the continuous quantum phase
transition and the thermal phase transition of the classical system expected
at any finite temperature. A coupling across blocks can also be achieved by
a weak global transverse field or other more complicated quantum fluctuations. 
Non-perturbative numerical results,
which layout the entire phase diagram in the presence of a transverse field,
are presented in the following section.

\section{\label{Sec3}Perturbations and Ashkin-Teller Criticality}

We now connect the different blocks using a weak perturbation which breaks
the conservation of $\sigma^z_i\tau^z_i$. 
In spin language this corresponds to a global transverse field,
leading to a Hamiltonian of the form 
\begin{equation}\label{pham}
\begin{split}
H=&\frac{-s}{2}\sum_{\langle i,j\rangle}\big(\sigma^z_i\sigma^z_j
+\tau^z_i\tau^z_j \big)\\
&-(1-s)\sum_i\big[p\sigma^x_i\tau^x_i+(1-p)(\sigma^x_i+\tau^x_i)\big].
\end{split}
\end{equation}
Here, the $\sigma^x(\tau^x)$ operator switches $00\to10\ (00\to01)$, 
effectively mixing blocks. In the weak perturbative limit
of $(1-p)\ll1$, this
can be seen as connecting blocks which have differing values for 
$\sigma^z_i\tau^z_i$ for only a few sites,
i.e those which have similarly sized chains in a similar arrangement. For
smaller $p$, blocks which have chains of substantially different sizes would
begin to couple as well, which would imply that the bound on the correlation
length would weaken as the system can now build in longer correlations through
a combination of blocks for the same value of $s$. In the
opposite limit of $p\to 0$, blocks are strongly coupled,
and the system can also be seen as two copies of transverse field
Ising models. This suggests that the system would undergo a continuous
transition, which would be in the Ising universality class of the appropriate
dimension. 

For $p=1$, the ground state sector is exactly a transverse field
Ising model on the appropriate lattice, as discussed in the previous section. 
In this limit, for all $s\in[0,1]$,
$M_P=\frac{1}{N}\sum_i \sigma^z_i\tau^z_i=\pm 1$ as 
all $\sigma^z_i\tau^z_i$ are either +1 or -1 for the
reference blocks. Here $P$ signifies polarization order, a term which is used
in literature discussing the Ashkin-Teller (AT) model \cite{ATintro,ATsource}, 
which is the higher dimensional
classical equivalent of our model. This will be discussed in more detail later
in this section.
For $s\to 1$, $\sigma^z$ and $\tau^z$ are each
disordered and with reducing $s$, they undergo an Ising transition where
they develop long range order. For $p<1$, at $s=0$ the paramagnet phase
has no long range order in fragment arrangements or either of the spin species
as the perturbation allows complete access to Hilbert space. The conditions
describe three phases, 1) complete paramagnetic phase with 
$\langle M^2_{P}\rangle=\langle M^2_{\sigma}\rangle
=\langle M^2_{\tau}\rangle=0$,
2) polarization ordering with $\langle M^2_{P}\rangle\neq 0,
\langle M^2_{\sigma}\rangle=\langle M^2_{\tau}\rangle=0$, and
3) ferromagnet with $\langle M^2_{P}\rangle\neq 0,
\langle M^2_{\sigma}\rangle=\langle M^2_{\tau}\rangle\neq 0$.

These three phases can also be understood in terms of the well-studied
AT model\cite{ATsource}. Our model of fluctuation-coupled Ising
systems can be mapped onto this classical model using the $d$-dimensional 
quantum to $d+1$-dimensional classical mapping based on the path integral
formalism. For a quantum system, the partition function, given by
Tr$[e^{-\beta H}]$, can be expanded in imaginary time using 
$\beta=n\Delta\tau$. This leads to a partition function of a classical model
in a higher dimension \cite{qtoc}, where $\sigma^x_i$ in the quantum model
is replaced by a bond $\sigma^z_i\sigma^z_{i+1}$, in the imaginary time 
direction. This substitution to Eq.~\eqref{pham} leads to an
anisotropic version of the AT model in $d+1$ dimensions.
The isotropic AT Hamiltonian is as follows:
\be\label{ATH}
H=-J\sum_{\langle i,j\rangle}\sigma^z_i\sigma^z_j
-J\sum_{\langle i,j\rangle}\tau^z_i\tau^z_j
-K\sum_{\langle i,j\rangle}\sigma^z_i\tau^z_i\sigma^z_j\tau^z_j.
\ee
For spin systems where there is no explicit coupling to the lattice, the
anisotropy is expected to be irrelevant (this equivalence is well-known for the 
simple transverse field Ising model).
The phase diagram of the above Hamiltonian in 2D as a function of 
$\frac{J}{T}$ and $\frac{K}{T}$ contains
the three phases shown by the 1D quantum Hamiltonian. 
The arguments presented until this point in this
section are valid for general lattices in all dimensions.

In 2D, the AT model is fairly well studied from a theoretical 
viewpoint \cite{ATsource}.
It was found that for $K>J$ and $J>0$, the system passes through two phase
transitions; from the paramagnet to the polarized phase and from the
polarized phase to the ferromagnet. Both these transitions are Ising-like as
a $Z_2$ symmetry is broken each time. At $K=J$, the polarized phase vanishes, 
and we have a direct transition from the paramagnet to the ferromagnet. The
universality class at this point is that of the $q=4$ Potts model in 2D. For
$0<K<J$, the system interpolates smoothly between two disconnected Ising models
($K=0$) and the $q=4$ Potts model. Along this interpolation, some of the 
critical exponents, such as the scaling dimensions of the polarization operator
and the energy density, vary smoothly \cite{Cardy}. 
This is expected as the energy
density coupling between the two species caused by the four spin term is
marginal in 2D and allows a smooth flow under a conformal field theory
description \cite{cCFT1}.
We check for a similar behavior in the coupled quantum Ising model on a
periodic chain, using stochastic series (SSE) expansion quantum Monte 
Carlo (QMC) \cite{SSE} as it
is a powerful and unbiased method of extracting thermodynamic expectation 
values for such systems. 

\begin{figure}[t]
\includegraphics[width=\hsize]{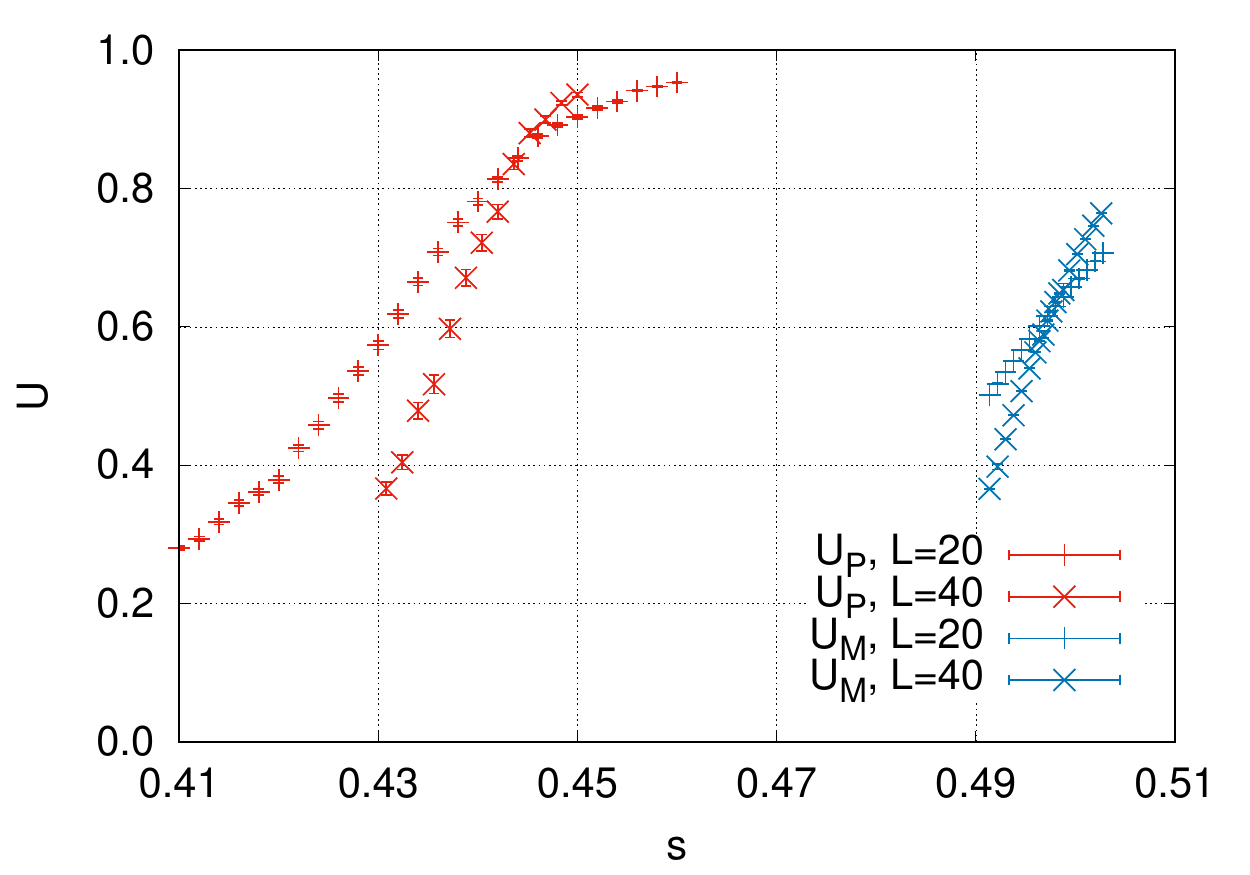}
\caption{Binder cumulant as a function of $s$ with crossing points for a pair
of sizes showing approximate locations of the two 
transitions at $p=0.95$.
(Inset) Extrapolation of crossing points of $(L,2L)$ for $U_M$ as a function of
$1/L$ fit to the form $f(x)=a+bx$ gives $s_c=0.497(1)$.}
\label{fdbs}
\end{figure}

\begin{figure}[t]
\includegraphics[width=\hsize]{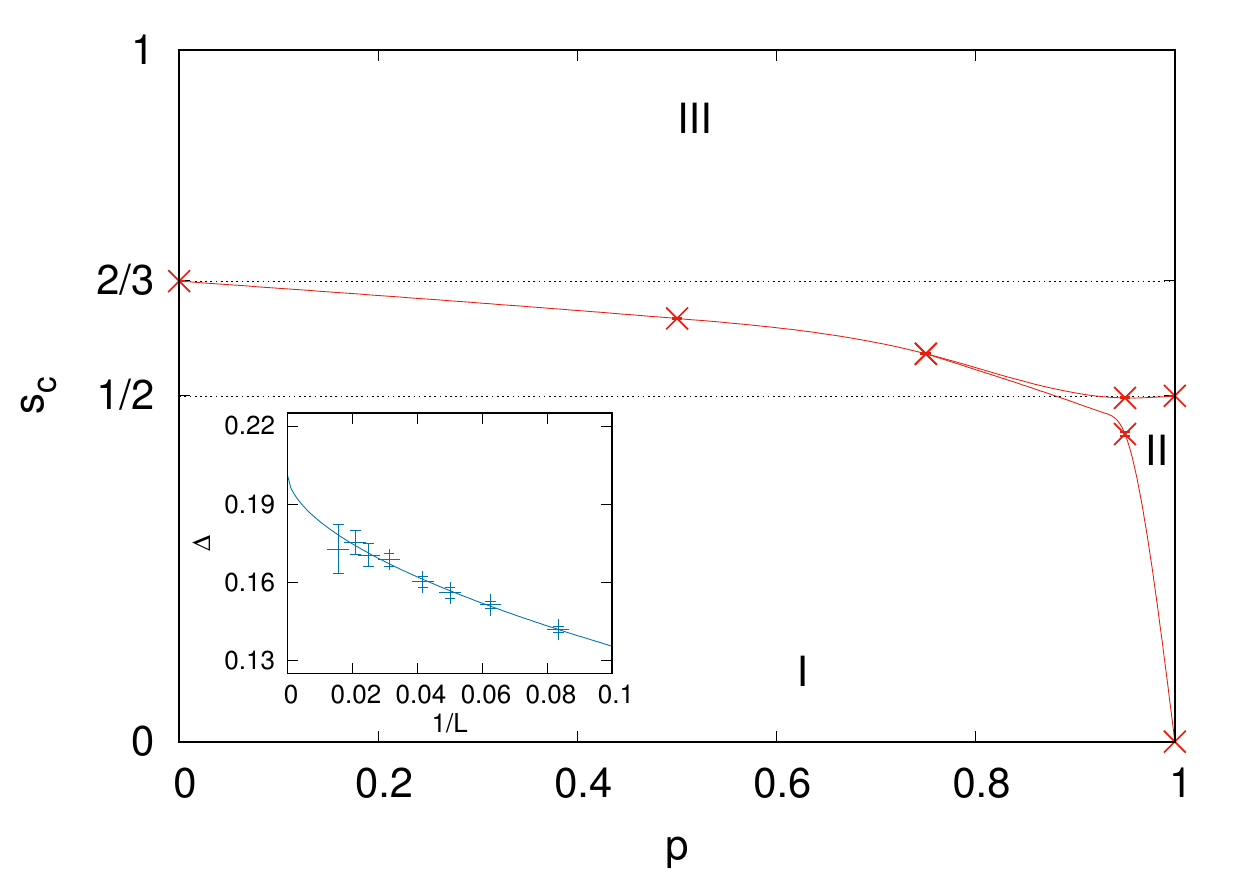}
\caption{Phase diagram of the model described by Eq.~\eqref{pham} with phases 
I. paramagnet, II. polarization ordered and III. ferromagnet with the AT line
of continuously varying exponents from $p=0$ to $p\approx0.75$. Inset:
Polarization exponent $\Delta_P(1/L)$ for $p=0.5$ extrapolated to
$\Delta_P(0)=0.20(1)$.}
\label{fphd}
\end{figure}

The $p=0$ limit corresponds
to the $K=0$ limit of the AT model and describes decoupled Ising models.
The $p=1$ limit has no paramagnetic phase and at some intermediate 
$p_{\textrm{ Potts}}$ , we would expect $q=4$ Potts criticality. 
For $p<p_{\textrm{ Pott}s}$, the system would 
trace out the line of continuously varying exponents and for 
$1>p>p_{\textrm{ Potts}}$, 
it would host all three phases along with two Ising transitions; one between 
the paramagnetic and polarized phases and the other between the polarized and
ferromagnetic phases. To investigate these
phase transitions, we define a Binder cumulant \cite{Binder} 
with coefficients corresponding to $Z_2$ symmetry breaking, as
\be\label{bcum}
U_M=\frac{3}{2}\bigg(1-\frac{1}{3}\frac{\braket{M^4}}{\braket{M^2}^2}\bigg),
\ee
where $M$ can denote either $M_P,M_{\sigma}$ or $M_{\tau}$. In the
regime where we have two Ising phase transitions, the
Binder cumulant is by this definition zero in the paramagnetic phase and unity
in the ordered phase, for whichever order parameter is considered. There is a
sharp transition in $U_M$ at the phase transition for large sizes and we need 
to study only one of $M_{\sigma}$ or $M_{\tau}$ as they are identical. By
tracking $U_P$ (corresponding to $M_P$) and $U_M$ (corresponding 
to $M_{\sigma}$), we notice two transitions for $p=0.95$
at distinct values of $s$. This is seen in Fig.~\ref{fdbs} through crossing
points of the Binder cumulant, and the inset shows an extrapolation of the
crossing points of $U_M$ as a function of inverse size, which leads to a
critical $s$ of $0.497(1)$ separating phases 2 and 3.
These two transitions are expected for values
of $p$ close to 1 until a point at which the $q=4$ Potts point is realized.
The scaling dimension of the spin operator is fixed at $\Delta_{\sigma}=1/8$
(which is the 2D Ising value)
along the critical line joining the $p=0$ and $p=p_{\textrm{ Potts}}$, 
whereas the polarization
operator has $\Delta_P=\Delta_{\sigma}+\Delta_{\tau}$ at the
decoupled point and $\Delta_P=\Delta_{\sigma}=\Delta_{\tau}$ at
the Potts point. The critical exponent $\nu$ varies from 1 (Ising value) to
3/2 (Potts value) along this line. From our simulations and finite size scaling
analysis following the method presented in Ref.~\onlinecite{Luck}, 
we observe that, at
$p=0.75$, $\nu=1.41(5)$ and $\Delta_P=0.13(1)$, indicating that
this point
is quite close to the Potts point (as can be seen in our approximate phase
diagram, Fig.~\ref{fphd}). The value of $\nu$ may be somewhat affected by 
logarithmic corrections expected in the exponents at the Potts point.
The same extrapolation at $p=0.50$
gives us $\nu=1.21(1)$ and $\Delta_P=0.20(1)$ (the extrapolation for $\Delta_P$
is shown in the inset of Fig.~\ref{fphd}), which are values
between the two extremes. This analysis shows us in a conclusive manner that
the system flows to the AT universality class in the thermodynamic limit.

\section{\label{Sec4}Relation to Pseudo-first Order Behavior}

The Binder cumulant is used in general to identify the nature of a phase
transition and the critical exponent $\nu$ for the correlation 
length (extracted from the slope).
Non-monotonic behavior in the Binder cumulant involving a minimum
is usually taken as a signature of a first-order transition, although this can 
only be confirmed by checking that the value of this negative peak diverges
as $L^d$ \cite{Binder}. A dip in the Binder cumulant had been misinterpreted 
to signal a first order transition \cite{FOE} 
for the frustrated $J_1$-$J_2$ classical 
Ising model on the square lattice where nearest
neighbors interact with a ferromagnetic bond of strength $J_1$ and next
nearest neighbors with an antiferromagnetic 
bond of strength $J_2$ \cite{JinSan}.In this model 
there exists a phase transition between a $Z_4$ symmetric striped
phase and a paramagnetic phase with increasing temperature. The dip was
taken to represent a first order transition until a detailed numerical study
by Jin \textit{et al.}
\cite{JinSan} showed that the cumulant dip
mapped onto the 2D $q=4$ classical Potts model, 
which also shows non-monotonicity
with a negative dip which does not diverge. The reason for
this behavior was traced to the shape of the distribution at the critical point
for these models \cite{Sannotes} and it was noticed that phase coexistence 
was not seen, which would be a characteristic of a first order transition.

\begin{figure}[t]
\includegraphics[width=\hsize]{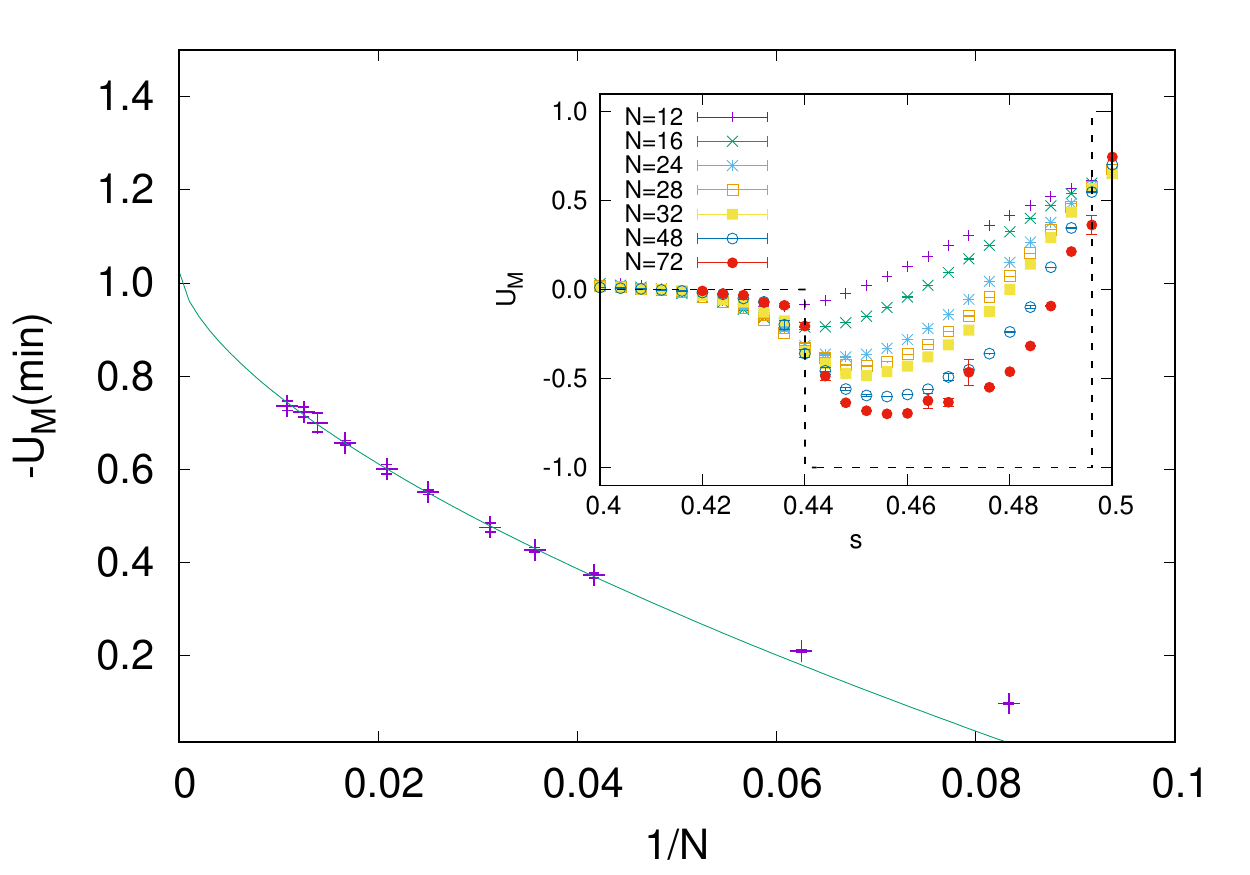}
\caption{The minimum value of $U_m$ for 1D coupled Ising chains, 
with Hamiltonian given by Eq.~\eqref{pham}, at $p=0.95$ as a function of 
length $L$ fit to the form 
$a+bL^{-c}$ converges to $1.02(3)$. Inset:
$U_M$ as a function of tuning parameter $s$ for various system sizes.
The thermodynamic from of $U_M$ is shown by the dashed line with transitions
at 0.44(1) and 0.497(1).}
\label{fbcum}
\end{figure}

Here we present the same kind of analysis
for our model of coupled Ising systems (Eq.~\eqref{oham}) in 1D
and argue that the negative peak
arises from an inappropriate 
definition of the Binder cumulant when investigating multiple phase
transitions. The Binder cumulant may evaluate to different values in different 
phases and if the phases are not well understood, this behavior can be
interpreted as arising from a first order transition. Even at special points 
such as the Potts point ($K=1$ point in the AT
model), which is known to harbor a continuous phase transition between trivial
paramagnetic and ferromagnetic phases, 
remnants of the polarization ordered phase cause non-monotonic behavior 
in the Binder cumulant. We will first show the non-monotonic behavior for the
coupled Ising chains, and will follow it up by explicitly showing the remnant
at the Potts point.

\newpage
\begin{widetext}

\begin{figure}[t]
\includegraphics[width=\hsize]{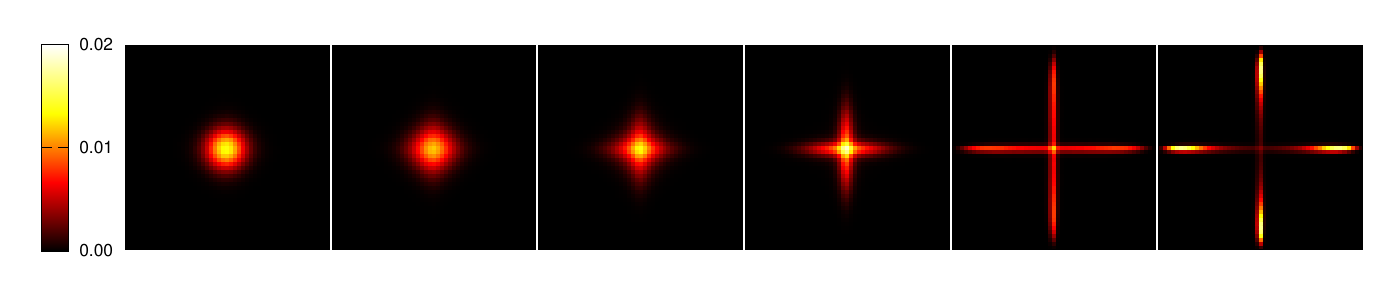}
\includegraphics[width=\hsize]{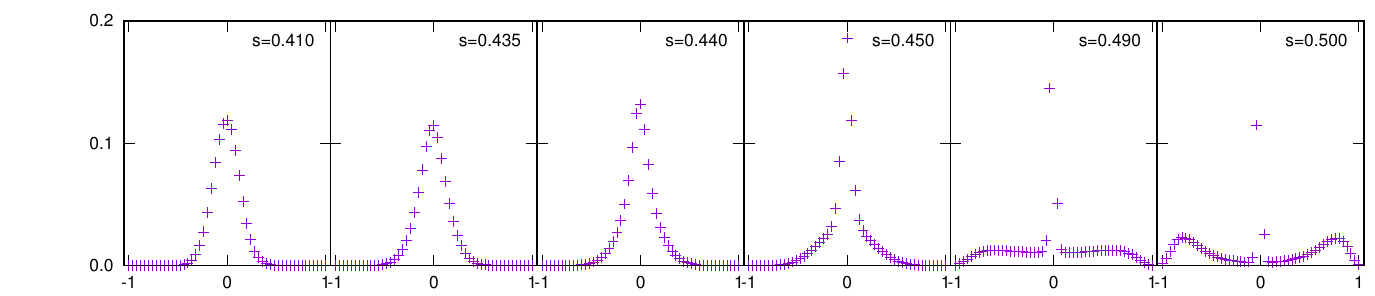}
\caption{Order parameter histogram as a function of the tuning parameter $s$.
Top: The 2D histogram of $M_x$ and $M_y$. Bottom: Marginal probability
distribution of $M_x$ at corresponding values of $s$.}
\label{fhist}
\end{figure}

\end{widetext}

If we consider the $p=0.95$ phase transitions presented in the previous
sections, we see that in the paramagnetic phase, the Binder cumulant can be
defined as
\begin{equation}\label{bcum2}
U_M=2-\frac{\braket{M^4}}{\braket{M^2}^2},
\end{equation}
instead of the definition used in Eq.~\eqref{bcum}, because the magnetization
can now be defined as a vector \textbf{\textit{M}}=$M_x \hat{x}+M_y \hat{y}$,
where $M_x (M_y)$ is the magnetization for the subset of spins with
$\sigma^z_i\tau^z_i=+1(-1)$.
This definition leads to $U_M=0$ for the paramagnetic phase and $U_M=1$ for
the ferromagnetic phase and is used for decoupled Ising systems as well as
systems with XY symmetry. Importantly, however, this definition of $U_M$ 
evaluates to $-1$ in the
polarization ordered phase as a global value of $\sigma^z_i\tau^z_i$ is chosen 
and only constrained
Ising like fluctuations are allowed along this axis forcing 
$\braket{M^4}$/$\braket{M^2}^2$=3, which can be calculated assuming
Gaussian probability distributions arising from the central limit theorem. 
If we use Eq.~\eqref{bcum2} for the
entire range of $s$ at $p=0.95$, in the thermodynamic limit, we would expect
a region where $U_M=0$, a region with $U_M=-1$ and a region with $U_M=1$.
A schematic of this is shown in the inset of Fig.~\ref{fbcum}.
For small sizes $U_M$ changes gradually and these values are not reached 
exactly. 

From Fig.~\ref{fdbs} and extrapolations similar to the one shown in its inset, 
we note that the paramagnetic to polarization ordered
transition occurs at $s=0.44(1)$ and the polarization ordered to ferromagnetic
one occurs at $s=0.497(1)$. Following the behavior of $U_M$ as defined above,
we find a non-monotonicity in the polarized phase where the dip extrapolates
to $-1$ (Fig.~\ref{fbcum}). We also study the histograms of the order parameter 
\textbf{\textit{M}} and clearly see the aligning of the 
polarization in Fig.~\ref{fhist}, where we present order parameter histograms 
for a 50-site system. We observe that the $(M_x,M_y)$ histograms look
substantially different from a continuous transition as the tails develop
a large distributed weight as we cross into the ferromagnetic phase,
even though the peak of the histogram is still at $(0,0)$, which 
represents the disordered phase. This can
be seen better in the marginal distribution of $M_x$ in the lower panels of
Fig.~\ref{fhist}, where we see that at $s\approx 0.49$, the histogram shows
a spread in weight outside of the disordered region, without a strong peak
at the order parameter value for the ordered phase.
This behavior is at odds with both a continuous phase transition, where one
must have a narrow peak which smoothly moves to $|M_x|=1$, and a first order
transition, where one must see two narrow peaks in the distribution but is
consistent with three phases. In the paramagnetic phase, the fluctuations
are Gaussian distributed in a radial pattern, whereas in the polarization
ordered phase, they are restricted to one dimensional distributions. In the
ferromagnetic phase, the system orders at one of the four peaks.

\begin{figure}[t]
\includegraphics[width=\hsize]{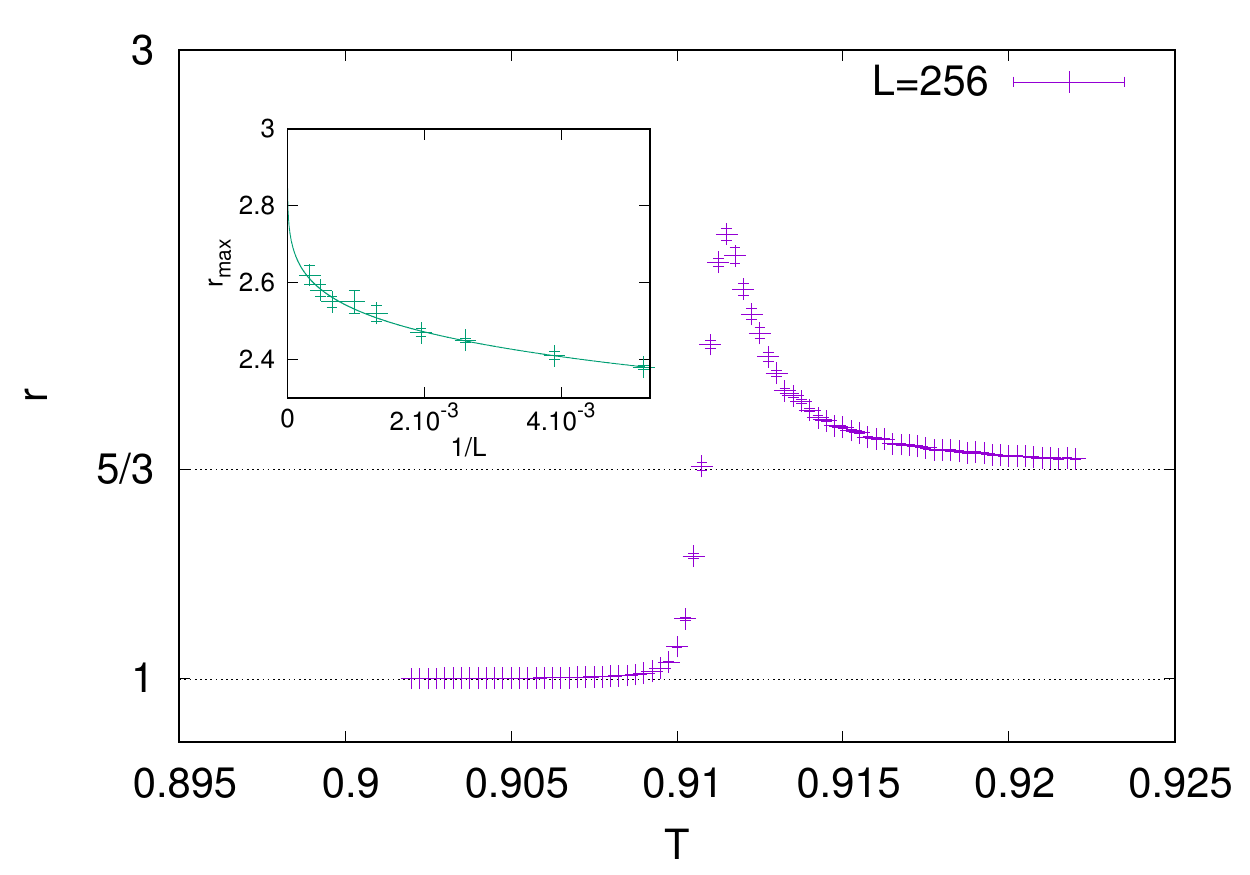}
\caption{Binder ratio for the 2D $q=4$ Potts model shows a peak at the
transition, as shown here for a 256$\times$256 system. Inset: Value of the
peak as a function of inverse linear size $1/L$, fit to a function of the 
from $f(x)=3-ax^b$.}
\label{fpot}
\end{figure}

These histograms are similar to those seen at the Potts point in the 
$J_1$-$J_2$ model. We have checked this in the more natural formulation of the
classical $q=4$ Potts model on a 2D square lattice, with a Hamiltonian given by
\be\label{PottsH}
H=-\sum_{\langle i,j\rangle}\delta_{q_i,q_j}
=-\sum_{\langle i,j\rangle}\cos(\theta_i-\theta_j),
\ee
where $q_i\in \{0,1,2,3\}$ are the possible states and which can be represented
as unit vectors forming a regular tetrahedron, implying the equivalence of the
two terms in Eq.~\eqref{PottsH} up to a global shift in the baseline for energy.
As mentioned above, if the fluctuations in the thermodynamic magnetization
are Ising like then $r=\braket{M^4}$/$\braket{M^2}^2=3$ and if they are
completely paramagnetic $r=5/3$, which can be seen by evaluating Gaussian 
integrals over the unit vectors chosen from a tetrahedron and which lie in 3D
space. In the ordered phase the fluctuations are small compared to the mean
and $r=1$. In the case of a typical continuous transition, $r$ would vary
monotonically from 1 to 5/3 from the ordered to paramagnetic phases. This is
not the case for the Potts model, as seen from our simulations in 
Fig.~\ref{fpot}, and we find a peak which grows for larger sizes. 
The peak appears to diverge logarithmically in the range which we have studied,
but we would expect this value to converge eventually (perhaps at $r=3$,
as shown in the inset of Fig.~\ref{fpot})
as we are studying a continuous phase transition. This
implies remnant effects of a polarization phase which cannot be explicitly
realized in this formulation of the Potts model. These effects persist up to
the largest lattice sizes (3072$\times$3072) we were able to study and may be
suppressed at even larger scales, in which case the origin of the new length 
scale would be of interest.

\section{\label{Sec5}Conclusions}

The coupled Ising model discussed here is a tractable system which can source
interesting dynamical behavior with excitations showing a restricted extent in
space.
Due to the intricate structure of 
non-interacting blocks which this system breaks into, curious features may be
manifest in the crossover between quantum and thermal phase transitions, and
we intend to study this in future work.
Upon the addition of perturbations it is expected that the system regains
ergodicity in a manner which depends on the particular perturbation used.
There has been a recent numerical study \cite{RN} which suggests that
long time scales persist even in the case of a 1D version of our model in the 
limit of weak global transverse fields creating a coupling across blocks.
In the presence of the same term, we have verified here that the system 
encodes a quantum realization of the AT model in a 
Hamiltonian made out of only two body terms explicitly for 1D and expect the
same in higher dimensions.

We have also identified a reason for pseudo-first order
behavior which is seen in the $q=4$ Potts model in 2D which corresponds to
a tricritical point with $q\le4$ corresponding to continuous transitions and
$q>4$ being first order transitions. This
could help explain the microscopic origin of the weak first order transitions
in the 1D quantum or 2D classical Potts model, which has been studied from
the perspective of complex conformal field theories \cite{cCFT1}. 
By switching off
the matrix element of the transverse field in the Potts model which connects
odd and even colors, all even color Potts models can be driven to exactly the
limit described here. 
The classical Potts model has also been independently 
studied in terms of restricted partitions \cite{PottsP}. 
Spin liquids with restricted dynamics have already been
found to have similar features \cite{Sha}, and we plan to develop a better
understanding for this in analogy with our model in future work.

\begin{acknowledgements}

We would like to thank Kedar Damle, Jun Takahashi, Anushya Chandran,
David J. Luitz and Rajesh Narayanan
for useful discussions. This work was supported by the NSF under Grant
No. DMR-1710170 and by a Simons Investigator Grant.
The computational
work was performed using the Shared Computing Cluster administered by 
Boston University's Research Computing Services. 
PP would like to thank Institute of Physics, 
Chinese Academy of Sciences for hosting a fruitful visit 
and facilitating collaboration.
We used QuSpin for checking the Quantum
Monte Carlo simulation against exact diagonalization calculations
\cite{Phil}.

\end{acknowledgements}


\begin{thebibliography}{999}

\bibitem{Senthil} T.~Senthil, A.~Vishwanath, L.~Balents, S.~Sachdev, and M.~P.~A.~Fisher, Science {\bf 303}, 1490 (2004).

\bibitem{Nahum} A.~Nahum, P.~Serna, J.~T.~Chalker, M.~Ortuno, and A.~M.~Somoza, Phys. Rev. Lett. {\bf 115}, 267203 (2015).

\bibitem{NahumX} C.~Wang, A.~Nahum, M.~A.~Metlitski, C.~Xu, and T.~Senthil, Phys. Rev. X {\bf 7}, 031051 (2017).

\bibitem{Bowen} B.~Zhao, P.~Weinberg, and A.~W.~Sandvik, Nature Physics {\bf 15}, 678-682 (2019).

\bibitem{qa1} S.~Boixo, Nature Physics {\bf 10}, 218 (2014).

\bibitem{qa2} E.~Farhi, D.~Gosset, I.~Hen, A.~W.~Sandvik, P.~Shor, A.~P.~Young,
and F.~Zamponi, Phys. Rev. A {\bf 86}, 052334 (2012).

\bibitem{qa3} E.~Farhi, J.~Goldstone, S.~Gutmann, J.~Lapan, A.~Lundgren, and
D.~Preda, Science {\bf 292}, 472 (2001).

\bibitem{IsingCFT} A.~A.~Belavin, A.~M.~Polyakov, and A.~B.~Zamolodchikov, Nucl. Phys. B {\bf 241}, 333 (1984).

\bibitem{PottsFT} F.~Y.~Wu, Rev. Mod. Phys. {\bf 54}, 235 (1982).

\bibitem{ClockFT} M.~Oshikawa, Phys. Rev. B {\bf 61}, 3430 (2000).

\bibitem{FracRev} R.~M.~Nandkishore and M.~Hermele, Annu. Rev.
Condens. Matter Phys. {\bf 10}, 295 (2019).

\bibitem{FracCham} C.~Chamon, Phys. Rev. Lett. {\bf 94}, 040402 (2005).

\bibitem{ETH1} P.~Sala, T.~Rakovszky, R.~Verresen, M.~Knap, and F.~Pollmann,
arXiv:1904.04266 (2019).

\bibitem{ETH2} V.~Khemani and R.~Nandkishore,
arXiv:1904.04815 (2019).

\bibitem{Sha} O.~Sikora, N.~Shannon, F.~Pollmann, K.~Penc, and P.~Fulde, Phys. Rev. B {\bf 84}, 115129 (2011).

\bibitem{ATintro} J.~Ashkin and E.~Teller, Phys. Rev. {\bf 64}, (5-6): 178-184 (1943).

\bibitem{Cardy} J.~L.~Cardy, J. Phys. A: Math. Gen. {\bf 20}, 13: L891 (1987).

\bibitem{ATsource} G.~Delfino and P.~Grinza, Nucl. Phys. B {\bf 682}, 521-550 (2004).

\bibitem{JinSan} S.~Jin, A.~Sen, and A.~W.~Sandvik, Phys. Rev. Lett. {\bf 108}, 045702 (2012).

\bibitem{ACCL} A.~Chandran and C.~R.~Laumann, Phys. Rev. B {\bf 92}, 024301 (2015).

\bibitem{Part1} G.~E.~Andrews, {\it Number Theory} (Dover Publications, Philadelphia, 1971).
	
\bibitem{Part2} C.~S\'andor, Integers {\bf 4}, A18 (2004). 
	
\bibitem{Perc1} H-O.~Heuer, J. Phys. A: Math. Gen. {\bf 26}, 333 (1993).

\bibitem{Perc2} T.~Senthil and S.~Sachdev, Phys. Rev. Lett. {\bf 77}, 5292
(1996).

\bibitem{Perc3} A.~W.~Sandvik, Phys. Rev. Lett. {\bf 96}, 207201 (2006).

\bibitem{Partm} G.~H.~Hardy, Ramanujan: {\it Twelve Lectures on Subjects
Suggested by His Life and Work}, 3rd ed. New York: Chelsea (1999).

\bibitem{YoungReiger} A.~P.~Young and H.~Rieger,
Phys. Rev. B {\bf 53}, 8486 (1996).

\bibitem{qtoc} M.~Suzuki, Comm. Math. Phys. {\bf 51}, 183 (1976).

\bibitem{cCFT1} V.~Gorbenko, S.~Rychkov, and B.~Zan, SciPost Phys. {\bf 5},
050 (2018).

\bibitem{SSE} A.~W.~Sandvik, Phys. Rev. B {\bf 59}, R14157 (1999).

\bibitem{Binder} K.~Binder, Z. Phys. B {\bf 43}, 119 (1981).

\bibitem{Luck} J.~M.~Luck, Phys. Rev. B {\bf 31}, 3069 (1985).

\bibitem{FOE} A.~Kalz, A.~Honecker, and M.~Moliner, Phys. Rev. B {\bf 84},
174407 (2011).

\bibitem{Sannotes} A.~W.~Sandvik, AIP Conference Proceedings {\bf 1297}, 1 (2010).

\bibitem{RN} Y.~Zhao, R.~Narayanan, and J.~Cho, arXiv:1906.12020 (2019).

\bibitem{PottsP} F.~Y.~Wu, G.~Rollet, H.~Y.~Huang, J.~M.~Maillard, C-K.~Hu, 
and C-N. Chen, Phys. Rev. Lett. {\bf 76}, 173 (1996).

\bibitem{Phil} P.~Weinberg and M.~Bukov, SciPost Physics {\bf 2}, 003 (2017).





\end{thebibliography}

\end{document}